\documentclass[12pt,draftcls,onecolumn]{IEEEtran}
\usepackage{amsmath}
\usepackage{amssymb}
\usepackage{amsfonts}
\usepackage{amscd}
\usepackage{mathrsfs}
\usepackage[final]{graphicx}
\usepackage{color}
\usepackage{url}
\usepackage{verbatim}
\newtheorem{theorem}{Theorem}

\newtheorem{defn}{Definition}

\newtheorem{construction}{Construction}

\begin{document}

\title{A Training based Distributed Non-Coherent Space-Time Coding Strategy}

\author{G. Susinder Rajan and B. Sundar Rajan
\thanks{G. Susinder Rajan and B. Sundar Rajan are with the Department of Electrical Communication Engineering, Indian Institute of Science, Bangalore-560012, India. Email:\{susinder,bsrajan\}@ece.iisc.ernet.in.}}

% avoiding spaces at the end of the author lines is not a problem with
% conference papers because we don't use \thanks or \IEEEmembership
%Homer Simpson\authorrefmark{2},
%Montgomery Scott\authorrefmark{3} and
%Eldon Tyrell\authorrefmark{4}}
%Telephone: (800) 555--1212, Fax: (888) 555--1212}
%\authorblockA{\authorrefmark{4}Tyrell Inc., 123 Replicant Street, Los Angeles, California 90210--4321}}

\maketitle

\begin{abstract}
Unitary space-time modulation is known to be an efficient means to communicate over non-coherent Multiple Input Multiple Output (MIMO) channels. In this letter, differential unitary space-time coding and non-coherent space-time coding for the training based approach of Kim and Tarokh are addressed. For this approach, necessary and sufficient conditions for multi-group decodability are derived in a simple way assuming a Generalized Likelihood Ratio Test receiver and a unitary codebook. Extending Kim and Tarokh's approach for colocated MIMO systems, a novel training based approach to distributed non-coherent space-time coding for wireless relay networks is proposed. An explicit construction of two-group decodable distributed non-coherent space-time codes achieving full cooperative diversity for all even number of relays is provided. 
\end{abstract}
%%%%%%%%%%%%Section 1 begins%%%%%%%%%%%%%%%%%%%%%%%%
\section{Introduction}
\label{sec1}

One of the efficient means to communicate over non-coherent MIMO (Multiple Input Multiple Output) channels is the training based non-coherent orthogonal designs approach of Kim and Tarokh \cite{KiT} which offers simple encoding, single complex symbol decoding along with full diversity. In this work, we generalize Kim and Tarokh's approach which result in multi-group decodable non-coherent space-time codes. Recently the authors of \cite{KiR,OgH2,JiJ} have proposed distributed differential space-time coding for wireless relay networks wherein all the terminals operate without the knowledge of any of the fading coefficients and yet achieve full cooperative diversity equal to the number of relays. However, the coding strategies proposed in \cite{KiR,OgH2,JiJ} put extra stringent conditions (as compared to the colocated MIMO case) on the unitary matrix codebook such as the existence of matrices that commute with all the codewords. This makes code constructions particularly difficult. For example, the distributed differential space-time code constructions in \cite{OgH2,JiJ} force all the codeword matrices to commute with each other.

The contributions of this letter can be summarized as follows.
\begin{itemize}
\item Generalization of the non-coherent orthogonal designs based construction of Kim and Tarokh \cite{KiT} by utilizing arbitrary linear designs instead of orthogonal designs alone. We refer to the resulting codes as training based non-coherent space-time codes. It is shown that by employing any full diversity coherent space-time code, a full diversity non-coherent space-time code can also be obtained.  
\item For the training based non-coherent space-time codes, necessary and sufficient conditions for multi-group decodability are derived in a simple and elegant manner assuming a Generalized Likelihood Ratio Test (GLRT) receiver and a unitary codebook. Moreover, the low complexity decoder for this case is described in a simple way. 
\item  Extending ideas from training based non-coherent space-time codes, a novel training based approach to distributed non-coherent space-time coding for wireless relay networks is proposed. This approach does not demand stringent conditions on the structure of the distributed space-time code such as commuting codewords which is the case for distributed differential space-time codes \cite{KiR,OgH2,JiJ}. Moreover, the channel coherence interval required for this strategy ($3R+1$ channel uses) is lesser compared to that required for distributed differential space-time coding \cite{KiR,OgH2,JiJ} ($4R$ channel uses). An explicit construction of two group decodable codes achieving full cooperative diversity for all even number of relays is also provided. 
\end{itemize}

The rest of the letter is organized as follows. In Section \ref{sec2}, we generalize the training based non-coherent space-time coding approach proposed in \cite{KiT}. A novel training based approach to distributed non-coherent space-time coding for wireless relay networks is proposed in Section \ref{sec3} and an explicit construction of $2$-group decodable codes achieving full cooperative diversity is also provided. Simulation results comprise Section \ref{sec4}.

\section{Multi-group decodable Training based Non-coherent Space-Time Coding}
\label{sec2}

In this section, we generalize the training based non-coherent orthogonal designs approach in \cite{KiT} by using arbitrary linear designs which are multi-group decodable instead of orthogonal designs alone. 

Let the number of transmit and receive antennas be denoted by $n$ and $m$ respectively. 
\begin{defn}\cite{RaR3}
\label{defn_lstbc}
A linear design $D(x_1,x_2,\dots,x_K)$ in $K$ real indeterminates or variables $x_1,x_2,\dots,x_K$ is a $n\times n$ matrix with entries being a complex linear combination of the variables. It can be written as $D(x_1,x_2,\dots,x_K)=\sum_{i=1}^{K}x_iA_i$ where, $A_i\in \mathbb C^{n\times n}$ are called the 'weight matrices'. A linear STBC $\mathscr{C}$ is a finite set of $n\times n$ complex matrices which can be obtained by taking a linear design $D(x_1,x_2,\dots,x_K)$ and specifying a signal set $\mathscr{A}\subset\mathbb{R}^{K}$ from which the information vector $X=\left[\begin{array}{cccc}x_1 & x_2 & \dots & x_K\end{array}\right]^T$ take values from, with the additional condition that $D(a)\neq D(a'), \forall\ a\neq a'\in\mathscr{A}$. A linear STBC $\mathscr{C}=\left\{D(X)|X\in \mathscr{A}\right\}$ is said to be $g$-group encodable (or $\frac{K}{g}$ real symbol encodable or $\frac{K}{2g}$ complex symbol encodable) if $g$ divides $K$ and if $\mathscr{A}=\mathscr{A}_1\times\mathscr{A}_2\times\dots\times\mathscr{A}_g$ where each $\mathscr{A}_i,i=1,\dots,g\subset\mathbb{R}^{\frac{K}{g}}$. An orthogonal design $\Theta(x_1,x_2,\dots,x_K)$ in $K$ real variables $x_1,x_2,\dots,x_K$ is a linear design satisfying $\Theta(x_1,x_2,\dots,x_K)^H\Theta(x_1,x_2,\dots,x_K)=\left(\sum_{i=1}^{K}x_i^2\right)I_n.$
\end{defn}

The proposed transmission matrix $S$ in \cite{KiT} has the form of $S=\left[\begin{array}{c}I_n\\ \Theta(x_1,x_2,\dots,x_K)\end{array}\right]$, where $\Theta(x_1,x_2,\dots,x_K)$ is an orthogonal design in $K$ real variables and of size $n\times n$. The portion corresponding to $I_n$ can be viewed as though pilots were sent from each of the transmit antennas. Hence this approach is referred to as training based. The authors of \cite{KiT} then propose to pair two real variables at a time to form $\frac{K}{2}$ complex variables and allow them to take values from a PSK constellation with appropriate number of points depending on the transmission rate that lead to single complex symbol decoding.

We propose to generalize this approach by replacing the orthogonal design $\Theta(x_1,x_2,\dots,x_K)$ by an arbitrary linear design $D(x_1,x_2,\dots,x_K)$. By doing so, we can still reap the benefits of low encoding complexity because a linear design has been utilized. To guarantee full diversity, we need to construct non-intersecting subspaces \cite{KiT,BeC}, i.e., the subspaces spanned by the columns of any two codewords should intersect trivially. To be precise, if $S_1$ and $S_2$ are two different codeword matrices then the matrix $\left[\begin{array}{cc}S_1 & S_2\end{array}\right]$ should have full rank equal to $2n$.

The codewords in our case are given by $S_1=\left[\begin{array}{c}I_n\\C_1\end{array}\right]$, $S_2=\left[\begin{array}{c}I_n\\C_2\end{array}\right],\dots$, $S_L=\left[\begin{array}{c}I_n\\C_L\end{array}\right]$, where $C_1,C_2,\dots,C_L$ are elements of a linear STBC $\mathscr{C}=\left\{D(X)|X\in \mathscr{A}\right\}$ as in Definition \ref{defn_lstbc}. For full diversity, we need the rank of $\left[\begin{array}{cc}I_n & I_n\\C_i & C_j \end{array}\right]$ to equal $2n$, which is same as $\mathrm{rank}\left(\left[\begin{array}{cc}I_n & 0\\C_i & C_j-C_i \end{array}\right]\right)=n+\mathrm{rank}\left(C_j-C_i\right)$. Thus if the matrix $\mathrm{C_i-C_j}$ has rank of $n$ for all $C_i\neq C_j\in\mathscr{C}$, full diversity is guaranteed. Thus, by simply vertically augmenting $I_n$ with a fully diverse linear STBC $\mathscr{C}$, we get a fully diverse training based non-coherent space-time code. 

Though, it is not necessary for the elements of $\mathscr{C}$ to be unitary matrices for achieving full diversity, we assume that they are unitary in the sequel in order to simplify the decoding algorithm. We assume the channel model to be as follows: $Y=SH+W$ where, $H$ denotes the $n\times m$ channel matrix, $S$ denotes the transmitted $2n\times n$ codeword, $Y$ denotes the $2n\times m$ received matrix and $W$ denotes the $2n\times m$ matrix with entries being a sample of zero mean complex Gaussian random variables with unit variance. We assume that the GLRT receiver is employed for detection at the receiver. For a unitary codebook, the GLRT receiver detects the codeword as follows \cite{BeC},
\begin{equation}
\hat{S}=\max_{i=1,\dots,L} \mathrm{Tr}\left[Y^HS_iS_i^HY\right].
\end{equation}
Let us simplify the above GLRT metric for codes of the specific form proposed. For our case $S_i=\left[\begin{array}{c}I_n\\C_i\end{array}\\\right]$ for some $C_i\in\mathscr{C}$. We have $\mathrm{Tr}\left[Y^HS_iS_i^HY\right]=\mathrm{Tr}\left[YY^HS_iS_i^H\right]$. Moreover $S_iS_i^H=\left[\begin{array}{cc}I_n & C_i^H\\C_i & C_iC_i^H\end{array}\right]=\left[\begin{array}{cc}I_n & C_i^H\\C_i & I_n\end{array}\right]$ where, the second equality is due to our unitary matrix assumption. Let us partition the received matrix $Y$ into sub-matrices $Y_1$ and $Y_2$ as follows: $Y=\left[\begin{array}{c}Y_1\\Y_2\end{array}\right]$. Let $Y_1$ denote the part of $Y$ corresponding to the transmission of $I_n$ (pilot part) and $Y_2$ denote the other part due to the encoded message. Then, we have $\mathrm{Tr}\left[YY^HS_iS_i^H\right]=\mathrm{Tr}\left[Y_1Y_1^H+Y_2Y_2^H\right]+ \mathrm{Tr}\left[Y_1Y_2^HC_i+Y_2Y_1^HC_i^H\right]$. The term $\mathrm{Tr}\left[Y_1Y_1^H+Y_2Y_2^H\right]$ does not depend on $C_i$ and hence can be ignored for decision purposes. Recall that $C_i$ was obtained by substituting for real variables $x_1,\dots,x_K$ in a linear design $D(x_1,\dots,x_K)$. Let $C_i=\sum_{j=1}^{K}x_j^iA_j$ where $x_1^i,x_2^i,\dots,x_K^i$ denote the specific values corresponding to $C_i$ taken by the variables $x_1,x_2,\dots,x_K$. Decoding to $C_i$ is thus same as decoding to the values taken by the set of variables or in other words decoding to the information vector $X$. Then the GLRT decoder can be rewritten as follows.
\begin{equation}
\label{eqn_GLRT_simplified}
\hat{X}=\left[\begin{array}{cccc}\hat{x}_1 & \hat{x}_2 & \dots & \hat{x}_K\end{array}\right]^T=\max_{i=1,\dots,L} \sum_{j=1}^{K}\mathrm{Tr}\left[Y_1Y_2^Hx_j^iA_j+Y_2Y_1^Hx_j^iA_j^H\right]
\end{equation}
It is clear that if $\mathscr{C}$ is $g$-group encodable, then the maximization in \eqref{eqn_GLRT_simplified} can be broken up into $g$ individual maximizations each of which is over only a subset of the $K$ variables since the real variables in a group takes values independently of the real variables in the other groups. Then the real variables in each group can be decoded independently of the real variables in the other groups. We refer to such codes as multi-group decodable codes. Note that the above decoder is very general in nature and also explains in a simple way how single complex symbol decoding can be done for the $2\times 2$ non-coherent orthogonal design proposed in \cite{KiT}.  

%%%%%%%%%%%%%%%%%%%%%%%%%%%%%%%%%%%%%%%%%%%%%%%%%%%%%%%%%%%%%%%%%%%%%
\section{A Distributed Non-coherent Space-Time Coding Strategy}
\label{sec3}

In this section, a novel training based approach to distributed non-coherent space-time coding for achieving cooperative diversity in wireless relay networks is proposed.

\begin{figure}[p]
\centering
\includegraphics{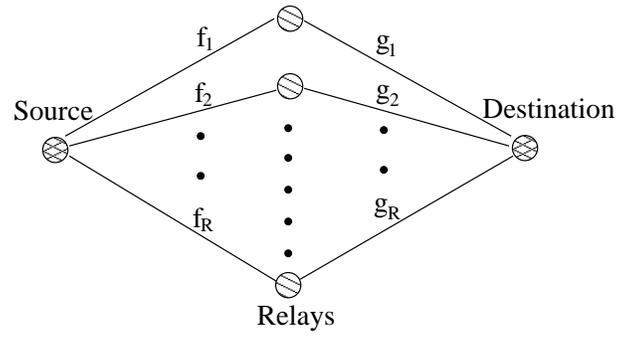}
\caption{A wireless relay network}
\label{fig_relay_pic}
\end{figure}

Consider a wireless relay network as shown in Fig. \ref{fig_relay_pic} with a source terminal, a destination terminal and $R$ relay nodes. We assume all the nodes in the network to be equipped only with single antennas. The fading gain of the channel between any two terminals is modeled by a zero mean complex Gaussian random variable with unit variance. The additive noise at all the terminals is modeled as a zero mean complex Gaussian random variable with unit variance. Let $f_i$ denote the channel fade coefficient between the source and the $i$-th relay and let $g_j$ denote the channel fade coefficient between the $j$-th relay and the destination. All the terminals are assumed to be symbol synchronized and half-duplex constrained. In this setting, we propose a training based distributed space-time coding strategy using which all the terminals can operate without the knowledge of any of the channel fading coefficients. It is important to note that though pilot signals are used in this strategy, none of the terminals are required to estimate the channel fade coefficients.

The transmission from source to destination consists of two stages. Each stage consists of two phases - a pilot phase and a communication phase. The first stage consists of $T_1+1$ channel uses. During the pilot phase of the first stage, the source transmits the complex number $1$ to all the $R$ relays using a fraction $\pi_1$ of the total power (sum of the power used by the source and all the $R$ relays) denoted by $P$. Then the received symbol at the $i$-th relay denoted by $r_i^p$ is given by $r_i^p=\sqrt{\pi_1P}f_i + n_i$ where, $n_i$ represents the additive noise at the $i$-th relay. During the communication phase of the first stage, the source transmits a vector $s$ of size $T_1\times 1$ taken from a codebook $\mathcal{C}$ satisfying $\mathrm{E}\left\{s^Hs\right\}=T_1$ using a fraction $\pi_1$ of the total power $P$ to all the $R$ relays. This vector $s$ actually carries the message intended to be communicated by the source to the destination. Thus the received vector during the communication phase at the $i$-th relay  denoted by $r_i^s$ is given by $r_i^s=\sqrt{\pi_1P}f_is+v_i$ where, $v_i$ represents the additive noise vector at the $i$-th relay.

\begin{figure}[p]
\centering
\includegraphics{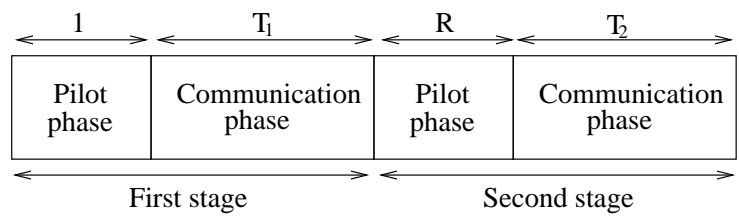}
\caption{Four phase transmission protocol}
\label{fig_protocol_phases}
\end{figure}

During the second stage, the relays linearly process the received signals from the source (which contains the pilots and the message) and relay the information to the destination. The pilot phase of the second stage consists of $R$ channel uses. Of these $R$ channel uses, one of them is allocated to each one of the $R$ relays for transmission. During its scheduled transmission slot, the $i$-th relay transmits a scaled version of $r_i^p$ using a fraction $\pi_2$ of the total power $P$. The symbol transmitted by the $i$-th relay is given by $t_i^p=\sqrt{\frac{\pi_2P}{\pi_1P+1}}r_i^p=\sqrt{\frac{\pi_1\pi_2P^2}{\pi_1P+1}}f_i+\sqrt{\frac{\pi_2P}{\pi_1P+1}}n_i$. In the communication phase of the second stage, all the $R$ relays transmit together a linearly transformed version of $r_i^s$ or its conjugate $r_i^{s^*}$ using a fraction $\pi_2$ of the total power $P$. For this purpose, each relay is equipped with a complex matrix $B_i$ of size $T_2\times T_1$, which we call the 'relay matrix' that satisfies $\parallel B_i\parallel_F^2\leq T_2$. The duration of the communication phase in the second stage is thus $T_2$ channel uses. To be precise, the vector transmitted by the $i$-th relay denoted as $t_i^s$ is given by $t_i^s=\sqrt{\frac{\pi_2P}{\pi_1P+1}}B_i\tilde{r}_i^s=\sqrt{\frac{\pi_1\pi_2P^2}{\pi_P+1}}f_iB_i\tilde{s}+\sqrt{\frac{\pi_2P}{\pi_1P+1}}B_i\tilde{v}_i$ where the notation $\tilde{x}$ denotes either $x$ or $x^*$ according to the context. The four phases in the entire transmission protocol are pictorially depicted in Fig. \ref{fig_protocol_phases}. The power allocation factors $\pi_1$ and $\pi_2$ have to be chosen so as to satisfy $\pi_1PT_1+\pi_2PRT_2=P(T_1+T_2)$.  Throughout this letter, we choose $\pi_1=1$ and $\pi_2=\frac{1}{R}$. In the proposed transmission protocol, the destination is scheduled to receive signals only during the second stage. Let $y_1$ and $y_2$ denote the received vector at the destination during the pilot phase and communication phase respectively of the second stage. Then, we have
$$
y_1=\sum_{i=1}^{R}g_it_i^p+u_1=\sqrt{\frac{\pi_1\pi_2P^2}{\pi_1P+1}}I_R\left[\begin{array}{c}\tilde{f}_1g_1\\ \tilde{f}_2g_2\\ \vdots\\ \tilde{f}_Rg_R\end{array}\right]+\sqrt{\frac{\pi_2P}{\pi_1P+1}}\left[\begin{array}{c}g_1n_1\\g_2n_2\\ \vdots\\ g_Rn_R\end{array}\right]+u_1
$$
where, the vector $u_1$ represents the additive receiver noise at the destination during the pilot phase of the second stage. Similarly, we have
$$
y_2=\sqrt{\frac{\pi_1\pi_2P^2}{\pi_1P+1}}\left[\begin{array}{cccc}B_1\tilde{s} & B_2\tilde{s} & \dots & B_R\tilde{s}\end{array}\right]\left[\begin{array}{c}\tilde{f}_1g_1\\ \tilde{f}_2g_2\\ \vdots\\ \tilde{f}_Rg_R\end{array}\right]+\left(\sqrt{\frac{\pi_2P}{\pi_1P+1}}\sum_{i=1}^{R}g_iB_i\tilde{v}_i\right)+u_2
$$
where, the vector $u_2$ represents the additive receiver noise at the destination during the communication phase of the second stage. Let $w_1=\sqrt{\frac{\pi_2P}{\pi_1P+1}}\left[\begin{array}{cccc}g_1n_1 & g_2n_2 & \dots & g_Rn_R\end{array}\right]^T+u_1$ and $w_2=\left(\sqrt{\frac{\pi_2P}{\pi_1P+1}}\sum_{i=1}^{R}g_iB_i\tilde{v}_i\right)+u_2$ which denote the equivalent noise as seen by the destination during the pilot and communication phases. Then we have the following signal model for the total received vector $y$ at the destination.
\begin{equation}
y=\left[\begin{array}{c}y_1\\y_2\end{array}\right]=\sqrt{\frac{\pi_1\pi_2P^2}{\pi_1P+1}}\left[\begin{array}{c}I_R\\
\begin{array}{cccc}B_1\tilde{s} & B_2\tilde{s} & \dots & B_R\tilde{s}\end{array}\end{array}\right]\left[\begin{array}{c}\tilde{f}_1g_1\\ \tilde{f}_2g_2\\ \vdots\\ \tilde{f}_Rg_R\end{array}\right]+\left[\begin{array}{c}w_1\\w_2\end{array}\right].
\end{equation}
Essentially we observe that the signal model becomes identical to a linear fading MIMO channel \mbox{$y=\sqrt{\frac{\pi_1\pi_2P^2}{\pi_1P+1}}SH+W$} where, $S=\left[\begin{array}{c}I_R\\
\begin{array}{cccc}B_1\tilde{s} & B_2\tilde{s} & \dots & B_R\tilde{s}\end{array}\end{array}\right]$, $H=\left[\begin{array}{cccc}\tilde{f}_1g_1 & \tilde{f}_2g_2 & \dots & \tilde{f}_Rg_R\end{array}\right]^T$ and $W=\left[\begin{array}{c}w_1\\w_2\end{array}\right]$. 
The difference here as compared to the case of colocated MIMO channels is that here the entries of the channel matrix $H$ are a product of two Gaussian random variables and the entries of the equivalent noise vector $W$ are not complex Gaussian distributed. 

For simplicity, we restrict ourselves to choosing $\mathcal{C}$ and the relay matrices $B_1,B_2,\dots,B_R$ such that the set of matrices $\mathscr{C}=\left\{\left[\begin{array}{cccc}B_1\tilde{s} & B_2\tilde{s} & \dots & B_R\tilde{s}\end{array}\right]\right\}$ consists of only unitary matrices. Let $|\mathscr{C}|=|\mathcal{C}|=L$ and let the elements of $\mathscr{C}$ be denoted by $C_1,C_2,\dots,C_L$. Then the distributed non-coherent space-time code consists of $L$ scaled unitary matrices of the form $S$ denoted by $S_1,S_2,\dots,S_L$. For such a unitary codebook, imitating the colocated MIMO case we propose to use a suboptimal mismatched decoder at the receiver as shown below: 
\begin{equation}
\label{mismatched_decoder}
\hat{S}=\max_{i=1,\dots,L} \mathrm{Tr}\left[Y^HS_iS_i^HY\right].
\end{equation}
We call this decoder as mismatched because the entries of the equivalent noise vector $W$ are not Gaussian distributed. Furthermore, this receiver is suboptimal because conditioned on knowing $g_j, j=1,\dots,R$, the covariance matrix of $W$ is a diagonal matrix and not a scaled identity matrix. This fact is not exploited by the decoder in \eqref{mismatched_decoder} and hence is suboptimal. The following theorem states that this suboptimal mismatched decoder already gives full cooperative diversity equal to $R$.

\begin{theorem}
\label{thm_criterion}
If $B_iB_i^H$ are diagonal matrices $\forall i=1,\dots,R$ and if $C_i^HC_i=C_iC_i^H=I_R,~\forall i=1,\dots,L$ then full diversity equal to $R$ is achieved by the suboptimal mismatched decoder in \eqref{mismatched_decoder} if $\mathrm{rank}\left(C_i-C_j\right)=R$ for all $C_i\neq C_j\in\mathscr{C}$.
\end{theorem}
\begin{proof}
The proof follows on the similar lines as the proofs in \cite{OgH2} and hence omitted.
\end{proof}

Observe that the sub-matrix of $S$ given by $\left[\begin{array}{cccc}B_1\tilde{s} & B_2\tilde{s} & \dots & B_R\tilde{s}\end{array}\right]$ can be viewed as a linear design if the vector $s$ is obtained from $T_1$ complex variables which take values from some signal set. Then the codewords of a distributed non-coherent space-time code look like $S_i=\left[\begin{array}{c}I_R\\C_i\end{array}\right]$, $C_i\in\mathscr{C}$ where $\mathscr{C}$ is now a linear space-time code. The difference here as compared to the colocated MIMO case is that $\mathscr{C}$ is obtained from a conjugate linear design (a linear design in which any column contains complex linear combinations of only the complex variables or only their conjugates) as opposed to any arbitrary linear design. In this correspondence we consider only such distributed non-coherent space-time codes since they are easier to study and their encoding complexity is also less. Moreover, the notion of multi-group decodable codes can then be utilized in the distributed setting also. 

Note that in the proposed coding strategy, the channels between all the terminals are assumed to be quasi-static for a duration of $T_1+T_2+R+1$ channel uses. Of the total $T_1+T_2+R+1$ channel uses, note that $R+1$ channel uses are employed for training purposes. Supposing the channel coherence interval is much more than $T_1+T_2+R+1$ channel uses, then we can stop the pilot phases after the first $T_1+T_2+R+1$ channel uses and the source can henceforth transmit data once very $T_1+T_2$ channel uses. In this work, we let $T_1=T_2=R$ for which the channel coherence interval should be $3R+1$ channel uses. At this juncture we would like to point out that distributed differential space-time coding \cite{KiR,OgH2,JiJ} on the contrary demands a channel coherence interval of $4R$ channel uses but can however enable the source transmit once every $2R$ channel uses always. Also note that the proposed strategy does not demand the existence of matrices that commute with the codeword matrices and a carefully chosen initial vector which is the case for distributed differential space-time coding \cite{KiR,OgH2,JiJ}. Furthermore though pilots have been employed in our transmission strategy, the relays do not estimate the fading gains from the source to the relays, but instead simply amplify and forward the pilots to the destination. 

\subsection{Explicit Coding}
In this subsection, we construct a class of $2$-group decodable fully diverse unitary space-time codes derived from PCIODs which can be employed as distributed non-coherent space-time codes. These codes can also be used in colocated MIMO systems for application either in the differential setup or in the training based setup as described in Section \ref{sec2}. PCIODs were first proposed for use as coherent distributed space-time codes in \cite{RaR2}. 
\begin{construction}\cite{RaR2}
Given an even number $R$, the rate one, $R\times R$ PCIOD $C_{P}$ is given as follows:

%%%%%%%%%%%%%%%%%%%%%%%%%%%%%%%%%%%%%%%%%%%%%%%%%%%%%%%%%%%%%%%%%%
{\scriptsize
\begin{equation}
\label{eqn_PCIOD_cons}
C_P =\mathrm{diag}\left\{\left[\begin{array}{cc}
x_1+ix_2 & -x_3+ix_4\\
x_3+ix_4 & x_1-ix_2\end{array}\right],\dots,\left[\begin{array}{cc}
x_{k}+ix_{k+1} & -x_{k+2}+ix_{k+3}\\
x_{k+2}+ix_{k+3} & x_{k}-ix_{k+1}\end{array}\right],\dots,\left[\begin{array}{cc}
x_{2R-3}+ix_{2R-2} & -x_{2R-1}+ix_{2R}\\
x_{2R-1}+ix_{2R} & x_{2R-3}-ix_{2R-2}
\end{array}\right]\right\}
\end{equation}
}
%%%%%%%%%%%%%%%%%%%%%%%%%%%%%%%%%%%%%%%%%%%%%%%%%%%%%%%%%%%%%%%%%%%%%%
There are totally $2R$ real variables in the linear design $C_{P}$.
\end{construction} 
We have $C_P^HC_P=\mathrm{diag}\left\{\left(\sum_{i=1}^{4}x_i^2\right)I_2,\dots,\left(\sum_{i=k}^{k+3}x_i^2\right)I_2,\dots\left(\sum_{i=2R-3}^{2R}x_i^2\right)I_2\right\}$ from which we infer that PCIODs do not lead to unitary codewords for arbitrary signal sets. But this can be accomplished by appropriately choosing multidimensional signal sets such that for all signal points \mbox{$\sum_{i=1}^{4}x_i^2=\dots=\sum_{i=2R-3}^{2R}x_i^2=1$}. To obtain full diversity, we first note that $\vert\Delta C_{P}^H\Delta C_{P}\vert=\left(\sum_{i=1}^{4}\Delta x_i^2\right)^2\dots\left(\sum_{i=2R-3}^{2R}\Delta x_i^2\right)^2$, where $\Delta C_P$ has been used to denote the difference matrix. Thus PCIODs do not offer full diversity for arbitrary signal sets. To get unitary matrices and full diversity we propose to choose the multidimensional signal points as follows. Firstly we form $R$ complex variables $s_1,s_2,\dots,s_R$ given by: $s_1=x_1+ix_2$, $s_2=x_3+ix_4,\dots,s_R=x_{2R-1}+ix_{2R}$. Now with this assignment of complex variables note that PCIODs are conjugate linear designs which is a necessary requirement for application in the distributed setting. We group these $R$ complex variables into two groups - First group:$s_1,s_3,\dots,s_{R-1}$ and Second group: $s_2,s_4,\dots,s_{R}$. Then $s_1$ and $s_2$ are allowed to take values independently from a PSK signal set with number of points depending on the transmission rate requirement. Then we let the complex variables $s_3,s_5,\dots,s_{R-1}$ to be some rotated versions of the specific value chosen by $s_1$. Similarly we let $s_4,s_6,\dots,s_{R}$ to be some rotated versions of the specific value chosen by $s_2$. Thus all the complex variables take values which lie on the unit circle. Then with this choice of multidimensional signal set, it is easy to check that the resulting codewords are fully diverse and unitary. PSK signal set has been employed in order to obtain unitary codewords. It is also clear that the resulting code is $2$-group encodable and $2$-group decodable. 
%%%%%%%%%%%%%%%%%%%%%%%%%%%%%%%%%%%%%%%%%%%%%%%%%%%%%%%
\section{Simulation Results}
\label{sec4}
\begin{figure}[p]
\centering
\includegraphics{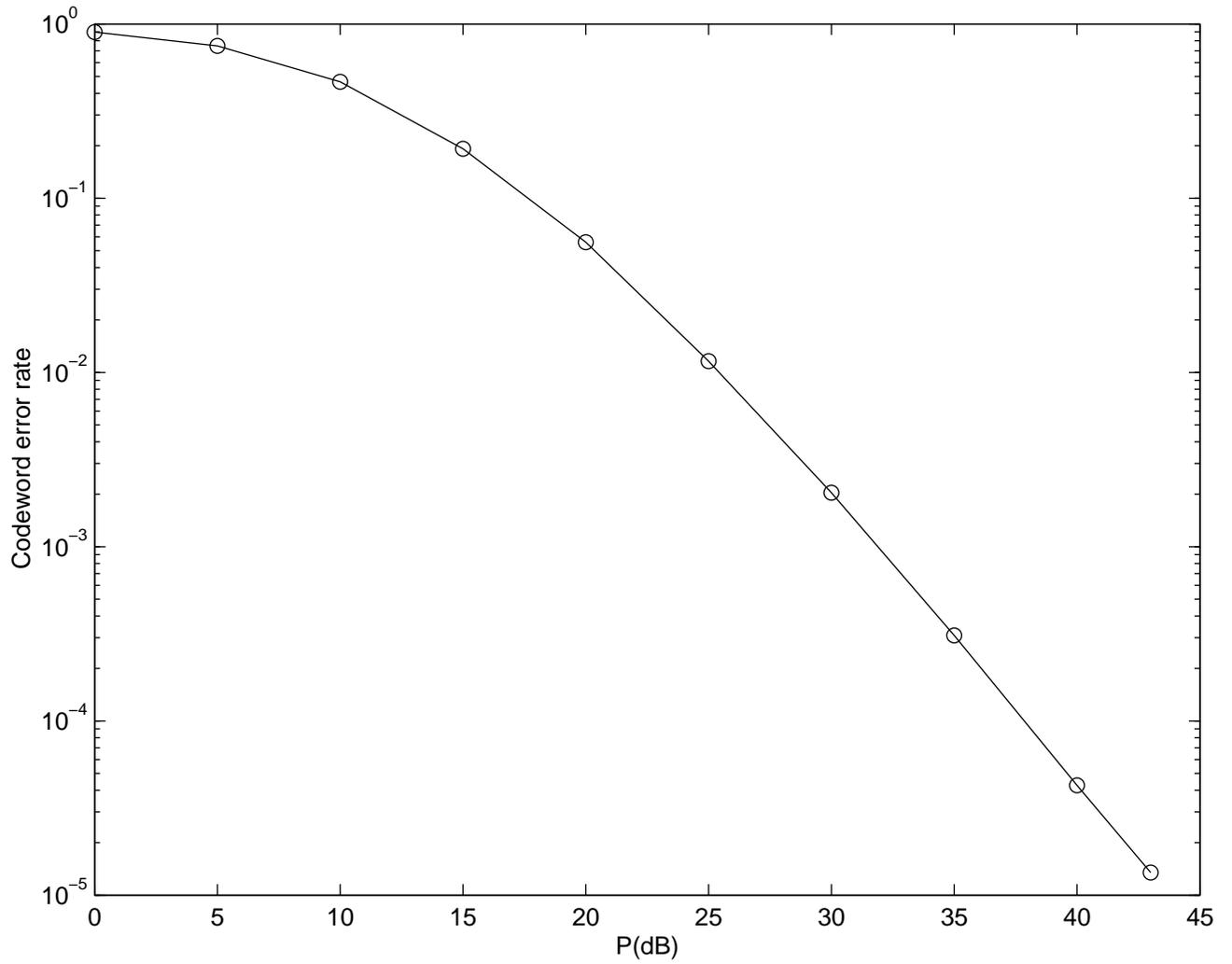}
\caption{Error performance of Alamouti design with QPSK constellation employed as a distributed non-coherent space-time code in a two relay system}
\label{fig_glrt_simulation}
\end{figure}

In this section, it is verified by simulations that full cooperative diversity is achieved if the rank criterion as stated in Theorem \ref{thm_criterion} is satisfied. The distributed non-coherent space-time code employed has been obtained using a unitary space-time code derived from the Alamouti design with a QPSK constellation. Thus $T_1=T_2=R=2$ and hence the transmission rate of the source is $\frac{2}{7}$ bits per channel use.  Fig. \ref{fig_glrt_simulation} shows the error performance of this code under single complex symbol decoding as in \eqref{eqn_GLRT_simplified} from which it can be observed that the slope of the codeword error rate is almost $2$ in the high SNR regime as expected.
%%%%%%%%%%%%%%%%%%%%%%%%%%%%%%%%%%%%%%%%%%%%%%%%%%%%%%%%%%%%%%%%%%%%%%%%%%%%%%%%
%\section*{Acknowledgement}
%The authors thank Yindi Jing and Hamid Jafarkhani for sending us a preprint of their recent works \cite{JiJ}.  
%%%%%%%%%%%%%%%%%%%%%%%%%

%%%%%%%%%%%%%%%%%%%%%%%%%%%%%%%%%%%%%%%%%%%%%%%%%
%%%%%%%%%%%%%%%%%%%%%%%%%%%%%%%%%%%%%%%%%%%%%%%%%%%
%%%%%%%%%%%%%%%%%%%%%%%%%%%%%%%%%%%%%%%%%%%%%%%%%%%%%%5
%%%%%%%%%%%%%%%%%%%%%%%%%%%%%%%%%%%%%%%%%%%%%%%%%%%

\end{document}